\documentclass[sigconf]{acmart}

\usepackage{flushend}
\usepackage{booktabs} % For formal tables
\usepackage{multicol}
\usepackage{multirow}
\usepackage[utf8]{inputenc}
\usepackage{array}
\usepackage[inline]{enumitem}
\usepackage{float}
\setlist[enumerate,1]{label=\textit{\alph*)}}
\copyrightyear{2018}
\acmYear{2018}
\setcopyright{acmcopyright}
\acmConference[MODELS '18]{ACM/IEEE 21th International Conference on
Model Driven Engineering Languages and Systems}{October 14--19,
2018}{Copenhagen, Denmark}
\acmBooktitle{ACM/IEEE 21th International Conference on Model Driven
Engineering Languages and Systems (MODELS '18), October 14--19, 2018,
Copenhagen, Denmark}
\acmPrice{15.00}
\acmDOI{10.1145/3239372.3239387}
\acmISBN{978-1-4503-4949-9/18/10}

\begin{document}
\title[Improving the Dev. Experience with Process Modelling Language]{Improving the Developer Experience with a Low-Code Process Modelling Language}
%\titlenote{Produces the permission block, and
%  copyright information}
%\subtitle{The Case of Outsystems}
%\subtitlenote{The full version of the author's guide is available as
%  \texttt{acmart.pdf} document}

\author{Henrique Henriques, Hugo Lourenço}
\affiliation{%
  \institution{OutSystems}
  \streetaddress{}
  \city{Linda-a-Velha} 
  \country{Portugal} 
}
\email{(henrique.henriques | hugo.lourenco)@outsystems.com}

\author{Vasco Amaral, Miguel Goul\~{a}o}
\affiliation{%
  \institution{NOVA LINCS, DI, FCT/UNL}
  \streetaddress{Universidade NOVA de Lisboa}
  \city{Lisboa} 
  \country{Portugal} 
}
\email{(vma | mgoul)@fct.unl.pt}

%\author{Miguel Goul\~{a}o}
%\affiliation{%
%  \institution{NOVA LINCS, DI, FCT/UNL}
%  \streetaddress{Universidade NOVA de Lisboa}
%  \city{Lisboa} 
%  \country{Portugal} 
%}
%\email{mgoul@fct.unl.pt}

% The default list of authors is too long for headers}
\renewcommand{\shortauthors}{H. Henriques et al.}

\begin{abstract}
	\textbf{Context: } The OutSystems Platform is a development environment composed of several  %domain-specific languages, 
	DSLs,
	used to specify, quickly build and validate web and mobile applications. The DSLs allow users to model different perspectives such as interfaces and data models, define custom business logic and construct process models. 
	\textbf{Problem: } The DSL for process modelling (Business Process Technology (BPT)), 
	%does not have the desired 
	has a low
	adoption rate and is perceived as having usability problems hampering its adoption. %is often used for solving problems out of the intended domain it was originally designed for. 
	This is problematic given the language maintenance costs.
	\textbf{Method: } We used a combination of %techniques, including 
	interviews, a critical review of BPT using the ``Physics of Notation'' and empirical evaluations of BPT using the System Usability Scale (SUS) and the NASA Task Load indeX (TLX), to develop a new version of BPT, taking these inputs and Outsystems' engineers culture into account. 
	\textbf{Results: } Evaluations conducted with 25 professional software engineers showed an increase of the semantic transparency on the new version, from 31\% to 69\%, an increase in the correctness of responses, from 51\% to 89\%, an increase in the SUS score, from 42.25 to 64.78, and a decrease of the TLX score, from 36.50 to 20.78. These differences were statistically significant.
%	\textbf{Limitations: }
	\textbf{Conclusions: } These results suggest the new version of BPT significantly improved the developer experience of the previous version. The end users background with OutSystems had a relevant impact on the final concrete syntax choices and achieved usability indicators.
\end{abstract}

%
% The code below should be generated by the tool at
% http://dl.acm.org/ccs.cfm
% Please copy and paste the code instead of the example below. 
%

 

%\copyrightyear{2018} 
%\acmYear{2018} 
%\setcopyright{acmcopyright}
%\acmConference[MODELS 2018]{ACM/IEEE 21th International Conference on Model Driven Engineering Languages and Systems}{October 14--19, 2018}{Copenhagen, Denmark}
%\acmBooktitle{MODELS 2018: SAC 2018: Symposium on Applied Computing , April 9--13, 2018, Pau, France}
%\acmPrice{15.00}
%\acmDOI{10.1145/3167132.3167264}
%\acmISBN{978-1-4503-5191-1/18/04}
\maketitle

\section{Introduction}
\label{sec:introduction}

Modelling Languages are increasingly adopted in industry. Improving the developer experience with those languages has a potential economic impact both by facilitating their adoption and by making developers more productive. Moody's seminal work on the ``Physics of Notations'' \cite{moody2009physics} has raised awareness to the importance of effective visual notations. However, there is scarce evidence and examples of industry-strength studies highlighting these benefits.

The \textit{OutSystems Platform} is used to create web and mobile applications with a set of integrated Domain-Specific Languages (DSLs). These DSLs are visual modelling languages that allow developing applications at a high abstraction level, hiding low-level details about creating and publishing those applications. This allows significantly faster development times and a higher quality result when compared to general-purpose languages \cite{outsystems2013outbynumbers}. The platform is used both internally and by external organizations, free-lancers, and even end-users, which develop their projects using this technology.

OutSystems includes a DSL called \textit{Business Process Technology} (BPT) for process modelling. BPT is used by developers with programming and process modelling knowledge and as a communication medium with business managers. Through interviews with OutSystems developers and data collected from recent projects, we found that BPT was not having the expected adoption rate (less targeted languages within OutSystems are used for process modelling) and was being used for purposes other than process modelling. Maintaining BPT has an associated cost. It was important to identify possible flaws in the language and make any necessary changes to raise its value for the company and its customers.

We used a combination of techniques for developing an improved version of BPT, including 
\begin{enumerate*}
    \item an analysis based on the ``Physics of Notations'' \cite{moody2009physics},
    \item interviews with professional BPT users,
    \item a ``crowdsourced'' approach to the production of an improved version of BPT's concrete syntax and
    \item its evaluation in terms of semantic transparency \cite{caire2013visual}, along with 
    \item a usability evaluation using the System Usability Scale (SUS) \cite{brooke1996sus} and
    \item a NASA Task Load indeX (NASA TLX) \cite{hart2006nasa} assessment of the effort involved in using BPT.
\end{enumerate*}

This combined methods approach has allowed for the production and implementation of a significantly improved version of BPT, in terms of its usability. The process is abstract enough to be applied to other visual modelling languages. The whole evaluation and improvement process of BPT was used as a testbed by OutSystems for the combination of these techniques to support language evolution.

We introduce BPT (section \ref{sec:BPT}), the language evaluation process for BPT (section \ref{sec:BPTevaluation}), and the new BPT proposal and the usability experiment (section \ref{sec:newBPTProposal}). On section \ref{sec:discussion} we discuss results and implications for practice. We then present related work (section \ref{sec:relatedWork}) and summarise the main conclusions (section \ref{sec:conclusion}). 
\section{Business Process Technology}
\label{sec:BPT}

\subsection{OutSystems Platform} 
\label{sec:platform}

\begin{figure}[ht]
	\centering
	\includegraphics[scale=0.45]{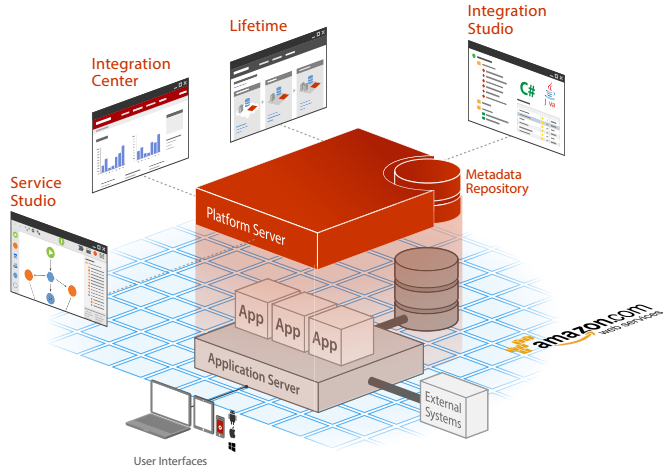}
	\caption{\textit{OutSystems Platform} architecture \cite{thesisbruno}.}
	\label{fig:architecture}
\end{figure}

The architecture \cite{lima2013architecture} is divided into three main components: \textit{Service Studio}, \textit{Platform Server} and \textit{Application Server} (Figure~\ref{fig:architecture}).

\vspace{-0.2cm}

\textit{Service Studio} is the development environment for all the DSLs supported by OutSystems. When the developer publishes an application, \textit{Service Studio} saves a document with the application's model and sends it to the \textit{Platform Server}. The IDE is divided into four main views: process modelling, interface flows, custom logic/APIs access and database modelling.

%% Stack deveria ser substituído por tool suite, ou outro termo mais adequado?
\textit{Platform Server} synthesises code given a particular stack (e.g., for a Windows Server \cite{winserver} using SQL Server \cite{sqlserver} this will be ASP.Net \cite{aspnet} and SQL code). The compiled application is then deployed to the \textit{Application Server}. The \textit{Platform Server} also includes a \textit{Scheduler Service}. This service manages the execution of steps within process models developed using BPT and also of scheduled jobs resulting from \textit{Timers}.

\textit{Application Server} runs on top of \textit{Oracle WebLogic} \cite{oracleweb}, \textit{JBOSS} \cite{jboss} or \textit{IIS} \cite{iss}. The server then stores and runs the developed application which is connected to a relational database management system, which can be \textit{SQL Server}, \textit{Oracle} \cite{oracle} or \textit{MySQL} \cite{mysql}. The SQL code generated by the \textit{Platform Server} is specific to the selected database management system.

%%%%%%%%%%%%%%%%%%%%%%%%%%%%%
%
%       End OutSystems
%
%%%%%%%%%%%%%%%%%%%%%%%%%%%%%

\subsection{Current BPT concrete syntax}
\begin{figure}[ht]
\vspace{-0.2cm}
    \centering
	\includegraphics[scale=0.45]{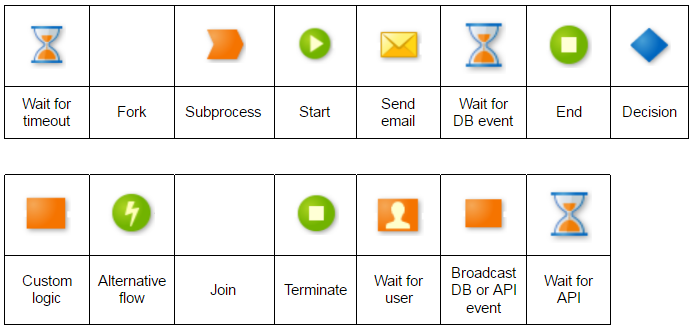}
	\caption{BPT development environment.}
	\label{fig:BPTOriginal}
\end{figure}

BPT users design, execute and manage processes which are fully integrated with applications built with the \textit{OutSystems Platform}. Figure \ref{fig:BPTOriginal} summarises BPT's concrete syntax.

\textbf{Start} starts the process flow. There can only be one Start in a process. An \textit{Alternative flow} is used to start a new parallel flow in the process. It has an attribute called \textit{Launch On} where the user defines what condition triggers the flow (e.g. a data-base event or an API call). \textit{End} and \textit{Terminate} share the same symbol, to either terminate the whole process or the particular flow it is connected to, respectively. \textit{Subprocess} calls another process. \textit{Wait for user} (also known as \textit{Human Activity}) is linked to a pre-developed \textit{Web Screen} (an Interface designed using another part of the \textit{OutSystems Platform}) and pauses the flow waiting for the user to trigger an action on that \textit{Web Screen}. An \textit{Automatic Activity} contains an action flow which is defined in a separate window. The action flow can include \textit{Custom Logic}, event broadcasts via the database (\textit{Broadcast DB}), or API calls (\textit{API Event}). \textit{Wait} pauses the process flow. The flow can then be resumed by a specific API call (\textit{Wait for API}), a database event (\textit{Wait for DB}) or an associated timeout (\textit{Wait for Timeout}). \textit{Send Email} is associated to a pre-developed email screen (which can contain dynamic data values). When the flow reaches this node it sends the email to the email addresses entered in the node's attribute. The \textit{Decision} node has \textit{n} outgoing flows and the chosen flow is decided based on custom logic defined in a separate window. %\textbf{Comment} allows users to write text in a small frame (the actual text is ignored by the compiler). 
Finally, although the language semantically supports parallelism, with a similar semantics of activity diagrams, there are no symbols for the concepts of \textit{Fork} and \textit{Join}.

\section{BPT Evaluation}
\label{sec:BPTevaluation}

\begin{figure}[ht]
	\centering
	\vspace{-0.2cm}
	\includegraphics[scale=0.47]{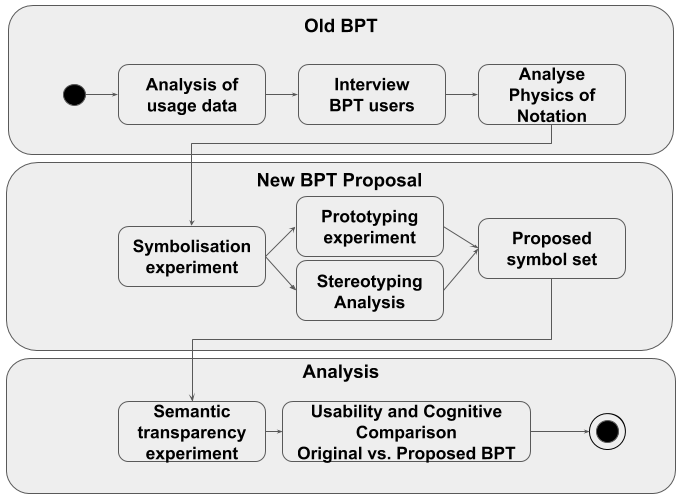}
	\caption{BPT evaluation process.}
	\label{fig:evalprocess}
\end{figure}

As depicted in Figure \ref{fig:evalprocess}, the process starts by evaluating the existing BPT, in order to identify improvement opportunities. The evaluation consists of an \textit{analysis of usage data from Outsystem's Platform logs},  \textit{interviews with BPT users}, an \textit{analysis of BPT using the ``Physics of Notations''}, and a \textit{usability analysis}.

\subsection{Analysis of usage data from the OutSystem's platform}

\noindent
\textbf{Procedure}
We analysed a repository of real application models (i.e. deployed in customers) and collected metrics concerning the percentage of those that actually use BPT, the percentage of BPT models that use process metadata (a bad smell in BPT, due to severe time performance implications), and the average number of nodes of BPT models.

\noindent
\textbf{Results}
From a repository of 5145 OutSystems application models available in the OutSystems platform, only 179 (around 3\%) used BPT. There were a total of 353 BPT models, of which 120 used process metadata. The 353 BPT models had an average of 18.7 nodes per BPT.  

\vspace{-0.2cm}
\subsection{Interviews with BPT users}
\label{subsec:BPTinterviews}

\noindent
\textbf{Procedure}
We interviewed 6 developers (2 Senior and 4 Lead Engineers) with at least 3 years of experience working with BPT developing applications for OutSystems clients. The interviews followed the Design Thinking philosophy of empathy interviews \cite{plattner2010design}, where the interviewee tells a story, from which one can collect more insights than those which would normally be available through answers to direct questions. The goal is to get to the root of problems by applying the \textit{five whys} technique \cite{dawson2012secrets}. This is an iterative interrogative technique to explore cause-and-effect relationships underlying a particular problem by repeatedly asking \textit{why} (at least \textit{five} times). The interviews covered the following topics:

\begin{itemize}
	\item \textbf{The context BPT is being used in.} Although BPT is a DSL it is possible that BPT is being used out of its domain. Usability issues can occur when a language is used in a domain that it was not designed for.
	\item \textbf{Why BPT was chosen.} This brings up the strengths of BPT and the features that the interviewee likes, and may help to start a conversation about things that can be further improved in said features.
	\item \textbf{What features are less used and why.} Features that are not getting much use may need to be changed, removed or better explained with training, documentation, etc. It is possible that the users do not use a certain feature because they do not know enough about it or its potential usefulness.
	\item \textbf{What features are missing.} With the daily usage of the language does the expert feel like there is something missing? Are there use-cases within the language's domain that can not be answered?
\end{itemize} 

\noindent
\textbf{Results}
\label{subsubsec:InterviewsResults}
The following insights about BPT were extracted from the interviews:
\begin{itemize}
    \item Development teams liked using BPT but were ``scared'' to do so due to low-level \textit{nuances}. They preferred to fall back to what they were used to (Timers).
    \item Parallelism was hard to model and so was identifying synchronization bugs.
    \item New team members could not start working with BPT without specialised training.
    \item Some clients explicitly requested the use of BPT.
    \item Maintaining a project with BPT was difficult and costly.
    \item BPT was often used outside of its domain. While BPT was designed to be a process modelling language it was also being used for event handling. These event handlers are very small processes (normally around three or four nodes). They start automatically in response to an event (like an API call), perform a small automatic action and then end. The problem is that the language runtime was not designed for this behaviour and, as such, does not perform well.
\end{itemize}

\vspace{-0.3cm}
\subsection{BPT analysis using the \textit{``Physics of Notations''}}
\label{subsubsec:BPTPoN}

\noindent
\textbf{Procedure}
We conducted a critical analysis of BPT, following the Physics of Notations (PoN) principles. We checked if BPT complies with each of the PoN nine principles, to determine the extent to which BPT's concrete syntax adheres to them and, in that process, identify concrete syntax improvement opportunities that would mitigate the identified non-conformities to those principles.

\noindent
\textbf{Results} 
In this section, we outline the main conclusions of our analysis of the BPT concrete syntax using the PoN as a reference framework. We do this by analysing each of the PoN 9 principles.

\textbf{Semiotic clarity:} 
There should be a one-to-one relationship between semantic constructs and concrete syntax. We found cases of \textit{symbol deficit}: while there is semantic support for using \textit{Forks} and \textit{Joins} to model parallelism, there were no special symbols for these. This symbol deficit may explain why our interviewees reported difficulties with identifying synchronization bugs and modelling parallelism (see section \ref{subsubsec:InterviewsResults}). We also found cases of \textit{symbol overload}: the \textit{Timeout} symbol is used for three types of \textit{waits} (all with different behaviours); the \textit{End} symbol is the same used for \textit{Terminate}; there are several ways of triggering a process but the \textit{Start} symbol does not reflect this. There are also some issues with platform consistency which may be regarded as a symbol overload problem when considering the other OutSystems DSLs. The \textit{Custom logic} and \textit{Broadcast DB or API event} constructs are represented by an orange ball in other OutSystems DSLs.

\textbf{Perceptual discriminability:}
Symbols with higher visual distance are easier to distinguish. No symbol had a visual distance greater than 2 visual variables. Colour and shape were the prevalent variables present in all the symbols. BPT uses textual differentiation, with all but the \textit{End} symbol having a label defined by the developer. The \textit{End} symbol has a label defining its behaviour: \textit{End} or \textit{Terminate}. There were no mechanisms for redundant coding.

\textbf{Semantic transparency:} 
Semantic transparency is the extent to which the meaning of a symbol can be inferred from its appearance. We made two complementary assessments of semantic transparency: a critical analysis of the concrete syntax, reported here, and an experimental evaluation, comparing the semantic transparency of the original BPT with the one of the proposed improved syntax of BPT (section \ref{sec:newBPTProposal}). In our critical analysis, we detected three visually opaque symbols. \textit{Call Subprocess} and \textit{Automatic Activity} have a shape that is not related to the corresponding notions. The \textit{Conditional start} symbol is represented with a lightning symbol which is not related to the notion of starting. However, this is mitigated by both \textit{Start} and \textit{Conditional start} sharing the same colour.

\textbf{Complexity management:} 
BPT has two mechanisms for dealing with complexity: the \textit{Call subprocess} construct and the \textit{Automatic Activity}. The subprocess construct calls another process and only proceeds after all flows of the subprocess finish. The automatic activity encapsulates logic but does not allow it to be reused. The \textit{Subprocess} construct is a good mechanism for managing complexity. It promotes reusable code and allows for several hierarchical levels. The \textit{Automatic Activity} could be improved by allowing re-usability.

\textbf{Cognitive Integration:}
A language should include some explicit mechanisms to support the integration of information from different diagrams. Two relevant mechanisms are \textit{Conceptual integration} and \textit{Perceptual integration}. Conceptual integration provides mechanisms (e.g. a summary diagram) to help the reader assemble information from separate diagrams into a coherent mental representation of the system. However, BPT does not support such mechanisms. Perceptual integration offers cues to simplify the navigation between diagrams. BPT supports \textit{orientation} and \textit{destination recognition} through the labelling of diagrams, but no explicit support for \textit{route choice} and \textit{monitoring}. 

\textbf{Visual expressiveness:}
The number of visual variables used in a notation defines its visual expressiveness. BPT has a visual expressiveness of 2. It only uses \textit{shape} and \textit{colour} as information carrying variables. \textit{Horizontal} and \textit{vertical position}, \textit{size}, \textit{brightness}, \textit{texture}, and \textit{orientation} are free variables in BPT.

\textbf{Dual coding:}
Textual encoding should supplement, rather than substitute graphics. However, BPT uses text as a way to distinguish symbols (e.g. the \textit{Custom logic} and the \textit{Broadcast DB or API event} are indistinguishable without textual labels). On a more positive note, BPT supports a text annotation construct, so that developers can add optional text to the process.

\textbf{Graphic economy:}
The number of symbols in the language should be manageable. BPT has 9 different symbols, which is over the recommended upper limit of 6 \cite{moody2009physics}. This shortcoming is mitigated by the fact that BPT is normally used within \textit{Service Studio}, where a toolbar also serves, in practice, as a key for the BPT diagrams. This mitigates the potential difficulty in remembering what each symbol means, which is in general more challenging for novices when understanding software engineering diagrams \cite{nordbotten1999effect}.

\textbf{Cognitive Fit:}
The cognitive fitness principle suggests that different representations of information are suitable for different tasks, audiences, and media. As is common in most Software Engineering languages, BPT uses a single visual representation for all purposes and audiences. However, this is not perceived as a significant shortcoming, as BPT’s notation is relatively small and simple to understand. Having separate dialects for experts and novices seems unlikely to bring significant benefits. Further research would be required to assess the potential impact of specific tasks on the usability of BPT. The notation is not easy to sketch since the language was designed to be used only within \textit{Service Studio}.

\vspace{-0.2 cm}
\section{New BPT proposal}
\label{sec:newBPTProposal}
\subsection{Research questions}
\label{subsec:researchQuestions}
During the analysis of the current version of BPT, we concluded that one of the areas that could be improved was its concrete syntax. Our goal was to create a new set of symbols that had a one to one relation between symbols and semantic constructs, and with a high level of semantic transparency. This was done by applying a modified method adapted from the work by Caire \textit{et al.} \cite{caire2013visual}. That said, the results of the Physics of Notations analysis should not be ignored and other factors (such as consistency) also needed to be considered. Three research questions guided our quasi-experiments on the semantic transparency of BPT:

\begin{itemize}
    \item \textbf{RQ1.} Is the original BPT concrete syntax semantically opaque?
    \item \textbf{RQ2.} Can participants unfamiliar with BPT design more semantically transparent symbols for BPT than the original?
    \item \textbf{RQ3.} Which concrete syntax (original, stereotype, prototype, proposed) is  more semantically transparent?
\end{itemize}

After conducting the semantic transparency evaluation of different versions of BPT, we further compared the \textit{usability} and \textit{cognitive effort} of the \textit{original BPT} with the BPT version proposed in this paper. This lead to three additional research questions:

\begin{itemize}
    \item \textbf{RQ4.} Which concrete syntax (\textit{original}, \textit{proposed}) leads to a better understandability of BPT models in the context of interpretation tasks?
    \item \textbf{RQ5.} Which concrete syntax (\textit{original}, \textit{proposed}) leads to a better understandability of BPT models, as perceived by practitioners after performing model interpretation tasks?
    \item \textbf{RQ6.} Which concrete syntax (\textit{original}, \textit{proposed}) leads to a lower cognitive effort, as perceived by practitioners while performing BPT model interpretation tasks?
\end{itemize}

\vspace{-0.3 cm}
\subsection{Research design}
\label{subsec:researchDesign}
The research design consisted of 6 related empirical studies, where the results of the earlier studies provide inputs to the later studies

We conducted five interrelated studies. Three were experiments involving novices. The participants used in each of the experiments were exclusive to that experiment, so that they would not be influenced by their own participation in other experiments. Additional material on all these empirical studies can be found in this paper's companion site \cite{CompanionSite}. %\footnote{\url{https://sites.google.com/fct.unl.pt/outsystems-bpt-evaluation/home}}.

\begin{enumerate}
	\item \textbf{symbolization experiment:} novices were asked to draw a set of symbols that they thought best represented the language constructs;
	\item \textbf{Stereotyping analysis:} a set of symbols was built based on \textit{the most common symbols drawn by the novices};
	\item \textbf{Prototyping experiment:} a different group of novices was asked to identify the best symbol for each construct - \textit{the most frequently selected symbol for each construct} was chosen;
	\item \textbf{Proposed symbol set:} a set of symbols was built, taking into account the results from the stereotyping and prototyping experiments, the interviews with users and the eye-tracking usability tests;
	\item \textbf{Semantic transparency experiment:} a third group of no-vices were asked to infer the meaning of each symbol. This was done for the \textit{original}, the \textit{stereotype}, the \textit{prototype} and the \textit{proposed} symbol sets. This last empirical study is the one where we finally evaluated the three research questions presented in section \ref{subsec:researchQuestions}.
	\item \textbf{Usability and cognitive effort comparison:} The original and the proposed BPT were compared in terms of their usability and of the cognitive effort associated in using them.
\end{enumerate}

\vspace{-0.2 cm}
\subsection{symbolization experiment}
\label{subsec:symbolizationExperiment}

\noindent
\textbf{i) Goal:}
Semantic transparency is achieved when users can infer the meaning of a symbol. A possible way to achieve an acceptable level of transparency is to have members of the target audience generate a set of symbols for the language. This was done applying the sign production technique \cite{howell1968population}, where novices were asked to draw symbols that best represent each of BPT's semantic constructs. 

\noindent
\textbf{ii) Participants and materials:}
The 24 participants were software developers (half of them were students, the other half professionals) between 18 and 32 years old. None of them had prior experience with BPT. The sign production questionnaire contains: 

\begin{itemize}
	\item a cover page with information about the study, a disclaimer and an out of context example to exemplify the expected answer format;
	\item fifteen questions (one for each of the semantic constructs) and respective answer box;
	\item a final page with a series of screening and demographic questions.
\end{itemize}

\noindent
\textbf{iii) Procedure:}
The participants received a printed questionnaire and were asked to answer them. They took from 30 to 40 minutes to complete the questionnaire. To process the questionnaires, we developed an application that receives as an input the questionnaire in digital format, cuts the answers into separate images, saves and indexes them. This allows viewing all the answers of a specific questionnaire or viewing all the answers for one specific construct. 
%The application also allows the user to upload the answers to the Usability Application. \par
%
%The decision to do the questionnaire in paper was based on previous studies and because it required the least amount of upfront work. However, given the difficulty to find participants (due to the required time investment) and the extra work to process the data we concluded that it would beneficial to be able to send the participant a digital version of the questionnaire and have them draw the answers using a stylus. This conclusion was reached while processing the data and therefore was not used in this study, but for future studies it would be best if both paper and digital options were available.

\noindent
\textbf{iv) Results:}
The outcome of this activity was a dataset of 24x15 symbol proposals for an improved BPT concrete syntax.

\subsection{Stereotyping analysis}
\label{sec:stereotypingAnalysis}

\noindent
\textbf{i) Goal:}
The stereotyping analysis builds on the assumption that, if several participants think of the same visual metaphor when proposing a representation for a given construct, this metaphor is likely to be easily recognizable by others. The goal of this task was to build a stereotype concrete syntax based on the most drawn metaphors for each construct.

\noindent
\textbf{ii) Materials:}
The input for this analysis was the symbol set collected from the questionnaires produced in the symbolization experiment, described in section \ref{subsec:symbolizationExperiment}.

\noindent
\textbf{iii) Procedure:}
The analysis of the drawings generated by the sign production technique was done using the judges' ranking method \cite{jones1983stereotypy}. The symbols were first classified into categories based on their conceptual similarity. Then, we chose the symbol that was most representative of the most frequent category.

For example, the symbols for the semantic construct \textit{Start process} were divided into five categories: \textit{media play button}, a \textit{text}, a \textit{power button}, a \textit{traffic light} and an \textit{on switch}. Of the total symbols, 11 were placed in the \textit{media play button} category, 6 in \textit{text}, 2 in \textit{power button}, 1 in \textit{traffic light} and 2 in \textit{on switch}. The remaining symbols were not categorised because they did not make sense or were unreadable. So, for the \textit{Start process} construct the chosen symbol was the symbol that best-represented \textit{media play button}. This process was repeated for each semantic construct.

\begin{figure}[ht]
\vspace{-0.2cm}
	\centering
	\includegraphics[scale=0.45]{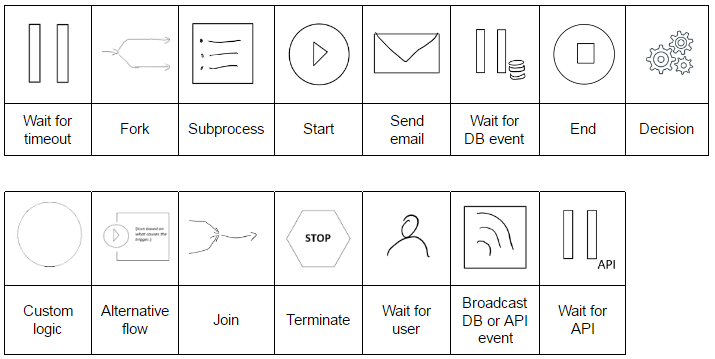}
	\caption{Set of stereotype symbols.}
	\label{fig:newstereotype}
\end{figure}

\noindent
\textbf{iv) Results:}
The stereotyping analysis resulted in a set of 15 symbols, one for each construct (Figure \ref{fig:newstereotype}). None of the symbols had an absolute majority. The large variety of symbols, and in some cases the lack of answer, illustrates the difficulty in creating a concrete representation for the constructs. 

\vspace{-0.3 cm}
\subsection{Prototyping experiment}
\label{subsec:prototyping}

\noindent
\textbf{i) Goal:}
A potential shortcoming of the stereotype symbol set is that the most drawn symbols are not necessarily those that better convey BPT's constructs. A visual metaphor may be a mnemonic of a construct's name but not a good representation of the concept itself \cite{caire2013visual,jones1983stereotypy}. As such, we conducted a prototyping experiment to identify which icons were better metaphors for the concepts, rather than for their names. 
\noindent
\textbf{ii) Participants and materials:}
We had a mixture of students and professionals, all software developers without previous knowledge of BPT, participating in the prototyping experiment. Most of the 16 participants were recruited through a digital third-party platform for usability tests called \textit{UsabilityHub} \cite{usabilityhub}. %\footnote{\url{https://usabilityhub.com}}. 

We created a questionnaire where each question had a description of a semantic construct and a set of possible symbols for that construct. The possible choices for each construct were a symbol from each category previously defined in \ref{sec:stereotypingAnalysis}. There were two versions of the questionnaire: one on paper (which can be found at the companion site) and another made available through \textit{UsabilityHub}.

\noindent
\textbf{iii) Procedure:}
Participants were asked to choose the best symbol to represent the description of the semantic construct. Most of the answers were collected through \textit{UsabilityHub}. The data exported from that platform was then complemented with the data collected from the few participants using the paper version of the questionnaire.

\begin{figure}[ht]
	\centering
	\includegraphics[scale=0.45]{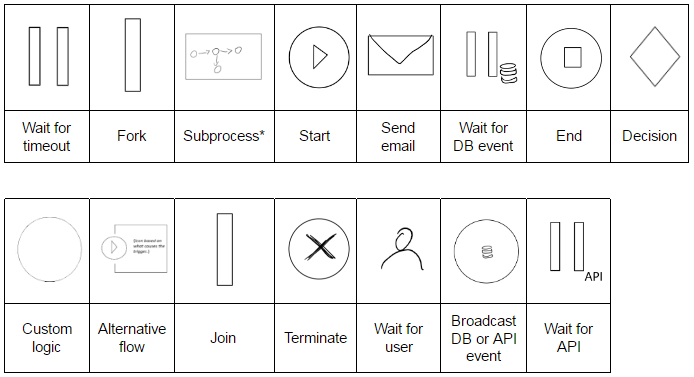}
	\caption{Set of prototype symbols.}
	\label{fig:newprototype}
\end{figure}

\noindent
\textbf{iv) Results:}
The set of most frequently chosen symbols for each construct is our \textit{prototype symbol set} (Figure \ref{fig:newprototype}). Again none of the symbols were selected by an absolute majority. While a few symbols matched the Stereotyping analysis, others were radically different. The \textit{Subprocess} construct is especially noteworthy since it has a dynamic effect: it is a real-time representation of the process being called which expands when the user mouse hovers it. 

\vspace{-0.2 cm}
\subsection{Proposed BPT}
\label{subsec:ProposedBPT}

\noindent
\textbf{i) Goal:}
A shortcoming of the stereotyping and prototyping experiments is that each symbol is generated independently. This neglects the development environment and the need for consistency with the rest of the OutSystems platform. Our goal was to propose an alternative BPT that would leverage this consistency concern, adding it to the knowledge gathered with the symbolization, stereotype and prototype experiments.

\noindent
\textbf{ii) Materials:}
We used as inputs the sets of stereotype and prototype symbols, as well as the feedback collected through interviews with OutSystems developers (already described in \ref{subsec:BPTinterviews}) and our knowledge of the rest of the OutSystems DSLs.

\noindent
\textbf{iii) Procedure:}
We proposed a new concrete syntax that takes into account consistency with the other OutSystems DSLs, while remaining as close as possible to the set of prototype symbols, as we expected these to be the most semantically transparent alternatives. However, while some changes were direct (simply changing the icon), others required changes to how the flows work and therefore go beyond the concrete syntax. This section goes over each of the changes made to BPT and how they were implemented. Note that, in the present stage, all these changes were developed as a prototype with the goal of testing the proposed concrete syntax. Further development is still necessary to make them part of the actual OutSystems platform.

%\noindent
\textbf{Adding new elements, updating symbols and syntactic ru-les.}
The original BPT set of symbols was comprised of nine symbols while the new one has a total of fifteen different symbols. So, in order not to increase the language's complexity by adding six new symbols to the toolbar, certain symbols were placed in groups and the symbol changes based on attribute values. As such, only two new symbols had to be added to the toolbar: \textit{Fork} and \textit{Join}.

The BPT language is defined by a meta-model which contains all the syntactical rules and constraints. This is consumed by a compiler which then generates a series of partial classes which can then be completed with the language's semantics. To add the new elements, nodes were created in the meta-model but their semantics were not touched at this stage given that the changes were just to prototype the syntax.

To view what causes a process to start, one has to view the properties of the process. During the usability tests, 100\% of the participants first looked for that information on the Start button. As such, the property was replicated to the Start button and any change made on button is updated in the process's property. This ensures there will not be any conflicts even though there is redundancy.

Some symbols have the same behaviour as the original ones. The only change made was the replacement of the original icon with the new one. This was the case for the \textit{Decision} and \textit{Wait} nodes. 

As the language now has explicit symbols for parallelism, the syntactical rules for outgoing arrows had to be updated. Originally any node could have \textit{N} outgoing and incoming arrows.  We changed this to ensure that only the \textit{Decision} (number of outgoing arrows is based on the condition) and \textit{Fork} can have \textit{N} outgoing arrows.

\begin{figure}[ht]
	\centering
	\includegraphics[scale=0.45]{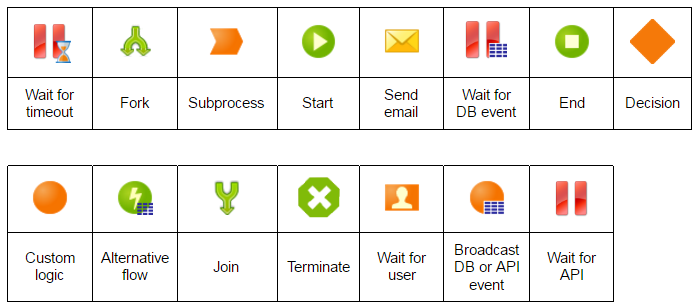}
	\caption{Set of proposed symbols.}
	\label{fig:newproposed}
\end{figure}

%\noindent
\textbf{Symbol groups.} 
When adequate, we grouped symbols together in order to reduce complexity. For each of these groups, we chose a symbol to represent the group. This symbol is then decorated with an overlay based on properties. We created the following groups:

\begin{itemize}
	\item \textbf{Waits.} This group includes the \textit{Wait for API call}, \textit{Wait for DB event} and the \textit{Wait for timeout} constructs. The symbol used to represent the group is the \textit{Wait for API call} symbol since it is the most generic of the three.
	\item \textbf{End and Terminate.} This group contains the \textit{End} and \textit{Terminate} symbols. They both have similar functionality but the \textit{End} is most commonly used, as such it was chosen to represent the group.
	\item \textbf{Start.} The launch of a process can be done in different ways. The default symbol for \textit{Start} is the one presented in \ref{fig:newproposed}. However, if the process is launched via a DB event then the symbol is updated with a small overlay.
	\item \textbf{Conditional start.} The proposed \textit{Conditional Start} symbol has an overlay representative of a DB event. But, the conditional start can also be triggered by an API call. As such, the default symbol used does not have an overlay but if the user chooses a DB event as a trigger then the overlay is applied.
\end{itemize}

%\noindent
\textbf{Using Actions in BPT.}
An Action is a piece of reusable code. There are different types of Actions: \textit{created by users} which can include any type of custom logic, a set of \textit{system actions} provided by \textit{Service Studio}, \textit{entity actions} that are used to manipulate the database, and \textit{API actions} which interact with triggers and APIs.

\begin{figure}[ht]
\vspace{-0.4cm}
	\includegraphics[width=1\linewidth]{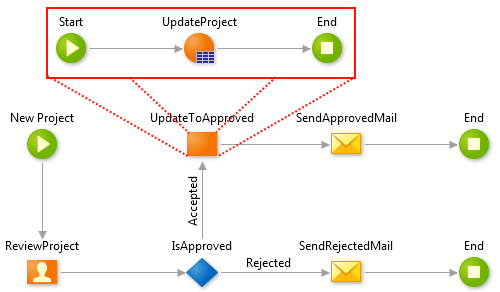}
	\caption{Automatic activity (original BPT)}
	\label{fig:oldauto}
\end{figure}

\begin{figure}[ht]
\vspace{-0.5cm}

	\includegraphics[width=1\linewidth]{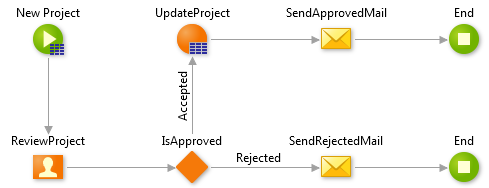}
	\caption{Actions (new BPT)}
	\label{fig:newauto}
\end{figure}

Originally in BPT, all action calls and logic had to be encapsulated in an \textit{Automatic Activity} (Figure \ref{fig:oldauto}). These were not reusable and at times created unnecessary complexity. Changes had to be made to ServiceStudio's syntax to allow Actions to be used in the BPT flow. With this, Actions have now replaced the \textit{Automatic Activity} (Figure \ref{fig:newauto}) since they provide generalization and re-usability while not creating unnecessary complexity.

\noindent
\textbf{v) Results:}
Figure \ref{fig:newproposed} presents the proposed concrete syntax for BPT.

The \textit{End} and \textit{Terminate} remain green (even though that could be considered semantically perverse) because \textit{End} is green on the rest of the platform. Changing it in the rest of the platform would have a large impact on established users so it was decided that it would be best to keep the green in these elements.

The most important changes when compared to the prototype symbol set concern the database related symbols (\textit{Wait for DB event} and \textit{Broadcast DB or API event}), the \textit{Wait for timeout} and \textit{Wait for API}, the \textit{Fork}, the \textit{Join} and the \textit{Alternative flow}:

\begin{itemize}
	\item \textbf{The database related symbols:} Both the stereotype and prototype symbol sets have use three cylinders to represent any activity related to a database. This is a common representation for databases. However, \textit{Service Studio} uses a different representation (a blue table). As such, all representations related to data were changed to maintain consistency.
	\item \textbf{Wait for timeout or API:} These were switched because waiting for an API call is an unconditional pause. A timeout is a condition. In order to be consistent with the database wait (which is also conditional), it was decided it would be best have an overlay on the conditions.
	\item \textbf{Fork and Join:} Arrows in ServiceStudio only specify flows and have no semantic definition. As such, the fork and join in the stereotype symbol would be a drastic change to the common behaviour of arrows. The metaphor from the prototype symbol set was not used because it makes it harder to differentiate forks from joins. The symbols proposed came in second on the prototyping experiment.
	\item \textbf{Alternative flow:} the symbols in both \textit{stereotype} and \textit{prototype} were too similar to the \textit{Start} symbol. We decided to maintain the current symbol and add an overlay depending on what triggers the alternative flow.
\end{itemize}

\subsection{Semantic transparency experiment}
\label{sub:trans}

\noindent
\textbf{i) Goal:}
The goal of this study was to \textbf{evaluate the semantic transparency of the 4 alternative symbol sets for BPT}: \textit{original BPT} (Figure \ref{fig:BPTOriginal}), \textit{stereotype BPT} (Figure \ref{fig:newstereotype}), \textit{prototype BPT} (Figure \ref{fig:newprototype}), and \textit{proposed BPT} (Figure \ref{fig:newproposed}). We conducted a blind interpretation study where participants had to infer the construct associated with each symbol. 
Comprehension tests \cite{zwaga1983evaluation} are commonly used to measure the symbol's transparency and is recommended by the International Organization for Standardization (ISO) for \textit{testing the comprehensibility of graphical symbols} \cite{ISO9186-1}.

\noindent
\textbf{ii) Participants and materials:}
The participants were 20 MSc students from Universidade Nova de Lisboa (UNL) from the Informatics course. None of them had previous knowledge of BPT. We created a survey for each of the 4 different sets of symbols. Each survey had a question for each of the symbols in each set, followed by a list with all of the language's semantic constructs. Participants were able to select one or more constructs that, in their opinion, were best represented by the corresponding symbol. The surveys were created in digital form. We developed a web application that asked the participant to input their (academic) email, and prevented answers from repeated emails. The application randomly redirected the participant to one of the four surveys, while keeping a balanced distribution of respondents among alternative notations (in practice, for every 4 participants, one would be randomly assigned to each of the alternative notations). This ensured that a tester could not answer more than one survey, to prevent bias. In the end, we had 5 respondents per notation. 

\noindent
\textbf{iii) Procedure:}
Each participants received a link to one of the 4 alternative questionnaires and filled it in. There was no fixed time limit for this task, but the estimated time for completion was \textit{``no longer than 15 minutes''}. 

\noindent
\textbf{iv) Hypotheses, parameters and variables:}
The independent variable was the concrete syntax (i.e. \textit{original}, \textit{stereotype}, \textit{prototype} or \textit{proposed} BPT). The dependent variable was the symbol's hit rate, used as an indicator of symbols comprehension (a proxy for semantic transparency). We hypothesized that the hit rate for the \textit{Original BPT} would be outperformed by all the alternatives, that the \textit{stereotype} would be outperformed by the \textit{prototype}, which, in turn, would be outperformed by the \textit{proposed BPT}:

\noindent
$Original BPT < Stereotype BPT < Prototype BPT < Proposed BPT$

\begin{figure}[!htb]
    \centering
    \includegraphics[scale=0.5, trim = 3cm 12.5cm 2.6cm 2.7cm, clip]{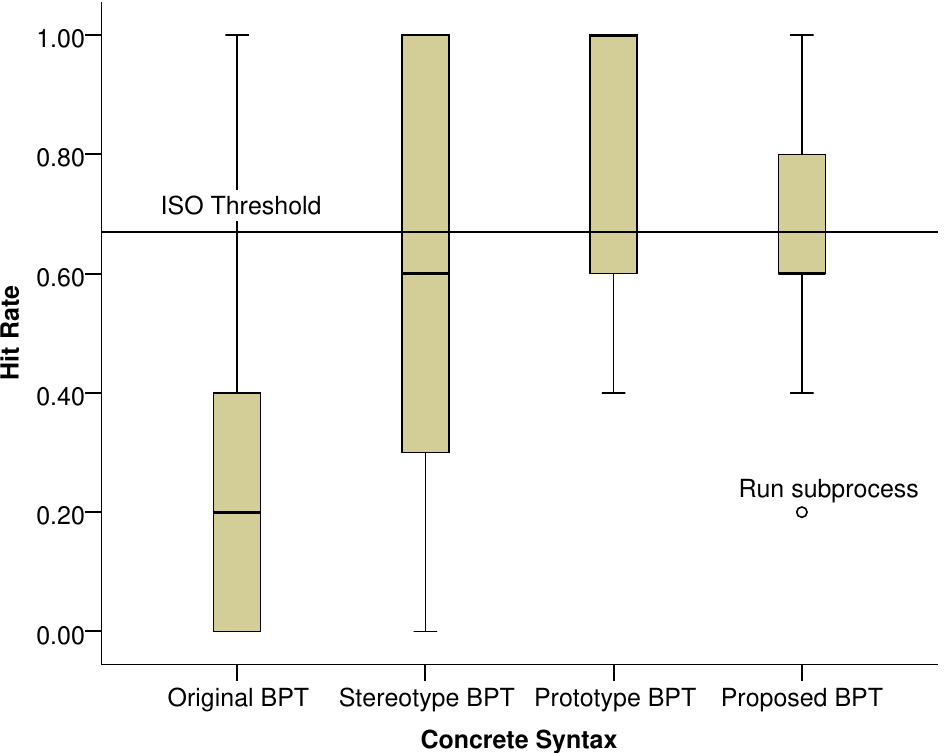}
    \caption{Hit Rate}\label{fig:hitRate}
\end{figure}

\noindent
\textbf{v) Results:} The results in Table \ref{tab:res} show that, on average, the \textit{Prototype} concrete syntax is the most semantically transparent set of symbols, with 79\% of hit ratio. The proposed BPT concrete syntax comes next, with around 69\% of hit ratio. Both are above the ISO threshold for comprehensibility (67\%) \cite{ISO9186-1}. However, only the \textit{Prototype BPT} has the mode above that threshold (Figure \ref{fig:hitRate}). Although with a similar median value, the Stereotype BPT had a lower mean hit rate, when compared to the \textit{Prototype} and the Proposed BPT. The \textit{Original BPT} obtained the lowest mean hit rate. In spite of not being the notation with the best hit rate, the \textit{proposed BPT} was selected in the end for further analysis, as it is the one that best fits into the remaining OutSystems DSLs landscape. As such further comparisons will focus on the original BPT and the \textit{proposed BPT}.

% Please add the following required packages to your document preamble:
% \usepackage{booktabs}
% \usepackage[normalem]{ulem}
% \useunder{\uline}{\ul}{}
\begin{table}[htb]
\centering
\footnotesize
\caption{Hit rate}
\label{tab:HitRate}
\begin{tabular}{@{}lrrrr@{}}

\toprule
\textbf{Construct} & \textbf{Original} & \textbf{Stereotype} & \textbf{Prototype} & \textbf{Proposed} \\ \midrule
\textbf{Wait for timeout} & {\underline{\textbf{1.00}}} & .60 & {\underline{\textbf{1.00}}} & \textbf{.80} \\
\textbf{Run subprocess} & .40 & .20 & \underline{.60} & .20 \\
\textbf{Start process} & .40 & {\underline{\textbf{1.00}}} & \textbf{.80} & .60 \\
\textbf{Send email} & \textbf{.80} & {\underline{\textbf{1.00}}} & {\underline{\textbf{1.00}}} & {\underline{\textbf{1.00}}} \\
\textbf{Wait for DB event} & .20 & {\underline{\textbf{1.00}}} & {\underline{\textbf{1.00}}} & \textbf{.80} \\
\textbf{End process} & .40 & .20 & {\underline{\textbf{1.00}}} & .60 \\
\textbf{Decision} & .40 & .0 & {\underline{.60}} & {\underline{.60}} \\
\textbf{Custom logic} & .0 & .0 & {\underline{.60}} & .40 \\
\textbf{Alternative flow} & .0 & .40 & .40 & {\underline{.60}} \\
\textbf{Fork} & .0 & {\underline{\textbf{1.00}}} & .40 & {\underline{\textbf{1.00}}} \\
\textbf{Join} & .0 & \textbf{.80} & .40 & {\underline{\textbf{1.00}}} \\
\textbf{Terminate} & .20 & {\underline{\textbf{1.00}}} & {\underline{\textbf{1.00}}} & \textbf{.80} \\
\textbf{Wait for user} & .60 & .60 & {\underline{\textbf{1.00}}} & \textbf{.80} \\
\textbf{Wait for API call} & .20 & .60 & {\underline{\textbf{1.00}}} & .60 \\
\textbf{Broadcast DB event} & .0 & \textbf{.80} & {\underline{\textbf{1.00}}} & .60 \\
\midrule
\textbf{Mean Hit Rate} & .31 & .61 & {\underline{\textbf{.79}}} & \textbf{.69} \\
\textbf{Standard Deviation} & .31 & .37 & .26 & .23 \\ \bottomrule
\end{tabular}
\end{table}

%\begin{figure}[ht]
%	\centering
%	\includegraphics[scale=0.65]{./images/results}
%	\caption{Semantic transparency results.}
%	\label{fig:semanresults}
%\end{figure}

%%%%%%%%%%%%%%%%%%%%%%%%%%%%%%%%
\vspace{-0.2 cm}
\subsection{Usability and cognitive effort comparison}
\label{subsec:BPTUsability}

\noindent
\textbf{i) Goal:}
Even if the language has a high transparency rating, assessed here indirectly through the \textit{hit rate}, it is important to conduct more usability experiments. The evaluations described in the previous sections evaluated symbols individually. This does not ensure a high language usability rating. We followed the method described in section \ref{subsec:BPTUsability} to further assess the \textit{proposed BPT} in terms of its usability and of the cognitive effort required to use it, contrasting it with the \textit{original BPT}.

\noindent
\textbf{ii) Participants and materials:}
A total of 25 subjects participated in this evaluation. They were professional software developers aged between 23 and 40 years old and experienced with the OutSystems platform. None of them had previous experience with BPT. 

We created two versions of the evaluation material, including three BPT models with varying levels of complexity that covered all the BPT language constructs. Semantically equivalent models were represented with the \textit{original BPT}, in the first version, and the \textit{proposed BPT}, in the second version. For each of those models, there was a set of questions, presented on a side window with a Google forms questionnaire. We asked participants to answer those questions, to assess their level of understanding about each model. The questions were objective (e.g. \textit{``What elements of the language interact with the database?''}) rather than subjective (e.g. \textit{``What do you think of X?''}), so that we could objectively assess whether the answer was correct or not. Overall, the questionnaires contained 9 questions. We also used a Simple Usability Scale (SUS) \cite{brooke1996sus} questionnaire and a NASA TLX questionnaire \cite{hart2006nasa} to assess the usability and cognitive effort required to answer the interpretation questions, respectively, as perceived by the participants.

The setup included a single-screen computer, equipped with an eye tracker, which was calibrated for each participant, at the beginning of the evaluation session. The OutSystems development environment was open, with a solution built with BPT.

\noindent
\textbf{iii) Procedure:}
We conducted one-on-one experiments with developers. 16 participants answered questions about models built with the \textit{original BPT}. Another group of 9 participants answered questions about the same models built with the proposed BPT. During the evaluation session, we recorded the contents of the screen, the eye tracking data of the participant while performing the tasks, and the voice of the participant. We encouraged the participant to follow a \textit{``think aloud''} protocol so that we would obtain richer data for analysis. After answering this questionnaire, the participant also answered a System Usability Scale (SUS) \cite{brooke1996sus} and a NASA TLX \cite{hart1988development} questionnaire. Finally, there was a short open discussion where the participant would talk about the issues he had with the language and, in most cases, suggested ideas for mitigating those issues.

\noindent
\textbf{iv) Hypotheses, parameters, and variables:}
The independent variable is the concrete syntax (\textit{Original BPT}, \textit{Proposed BPT}). The dependent variables are the answer correction rate, the SUS score, and the NASA TLX score. For these three variables, we hypothesized that the \textit{Propose BPT} would outperform the \textit{Original BPT} in terms of understanding tasks with BPT models, the perceived usability of BPT and the perceived cognitive effort spent while using it. 

\noindent
\textbf{v) Results:}
\label{subsec:UsabilityResults}
%Figure \ref{fig:graphan} contains a graphic overview of the results.
%
%\begin{figure}[ht]
%	\centering
%	\includegraphics[scale=0.45]{./images/graph}
%	\caption{Graph of answer distribution}
%\label{fig:graphan}
%\end{figure}
%
%Each number present in Figure \ref{fig:graphan} corresponds to a question which can be found on table \ref{tab:res}. Said table also includes a more detailed overview of the results and comments about the New BPT tests.
Table \ref{tab:res} presents the success rate for each question, for \textit{Original BPT} and \textit{Proposed BPT}. The right column presents a short comment concerning how the change of the concrete syntax affected the corresponding answer, as perceived from the observation of the participation of our subjects in the usability experiment (including the eye tracking data) and from their own feedback.

{\footnotesize \begin{table*}[!htb]
	\centering
	\caption{Results of usability tests}
	\label{tab:res}
\begin{tabular}{llll}
		\toprule
		& \multicolumn{2}{l}{Correct Answer Rate (\%)}                   &                   \\ \cmidrule(lr){2-3}
		Question                                              & \multicolumn{1}{l}{Original} & \multicolumn{1}{l}{Proposed} & Proposed BPT Comments                                                                                                                                                                                                                             \\ \midrule
		\textbf{1.} What causes the process to start?                  & 6.25                             & 100                         &  Participants had no problem to find the information since it is now present on the Start \\
		&&& button (which was the first place to be checked).\\
		\textbf{2.} Which nodes require human interaction?             & 100                              & 100                         & Same result as the Original BPT test. This was expected since the symbol was not changed.                                                                                                                                                  \\
		\textbf{3.} What nodes send e-mails?                           & 100                              & 100                         & This is the same case as the question above.                                                                                                                                                                                           \\
		\textbf{4.} Can this process fail?                             & 62.5                             & 100                         & Testers had a much easier time understanding the difference between \textit{End} and \textit{Terminate}.                                                                                                                                                 \\
		\textbf{5.} Why can this process fail?                          & 31.25                            & 66.66                       & There is still some confusion about the scope of Terminate. It is not clear if the Terminate \\
		&&&in the sub-process also kills the parent.\\
		\textbf{6.} What node finishes a process flow?                 & 31.25                            & 100                         & With different symbols for each construct testers no longer confuse the two.           \\
		\textbf{7.} What node finishes all process flows?              & 31.25                            & 100                         & Same as the above.                          \\
		\textbf{8.} Who are the participants (actors) in this process? & 0                                & 33.33                       & No changes were made to this node so identifying who interacts with the process is still \\
		&&&a problem.\\
		\textbf{9.} Who is responsible for each node?                  & 100                              & 100                         & The testers matched easily the actors (due to the intuitive labels on the nodes). Without \\
		&&&labels the results would be much worse. \\ \bottomrule
	\end{tabular}
\end{table*}
}

%To determine the language's usability we analysed the SUS and TLX scores gathered from usability experiments. By analysing these two scores we can determine whether the difference between the original and the new BPT language are significant and relevant. As a reminder, a higher SUS score is better while a lower TLX score is better.

Table \ref{tab:desc} contains descriptive statistics for the SUS and TLX data collected while performing the usability tests. The table is grouped by score type and each type contains statistics for the original BPT language (BPT) and the new BPT language (New BPT).

{\footnotesize
\begin{table}[ht]
	\centering
	\caption{Descriptive statistics}
	\label{tab:desc}
	\begin{tabular}{@{}llrrrrrr@{}}
		\toprule
		\multicolumn{1}{l}{} & \multicolumn{1}{c}{Language} & \multicolumn{1}{c}{N} & \multicolumn{1}{c}{Mean} & \multicolumn{1}{c}{Std. Dev.} & \multicolumn{1}{c}{Skew.} & \multicolumn{1}{c}{Kurt.} & \multicolumn{1}{c}{S-W} \\ \midrule
		Correct & Original BPT & 9 & 0.51 & 0.40 & 0.227 & -1.739 & 0.085\\
		& Proposed BPT & 9 & 0.89 & 0.24 & -2.121 & 4.001 & 0.000\\
		SUS                  & Original BPT                          & 16                    & 42.25                    & 11.72                       & -0.021                   & -0.67                    & 0.896                            \\
		& Proposed BPT                      & 9                     & 64.78                    & 10.02                       & 0.213                    & -1.05                    & 0.697                            \\
		TLX                  & Original BPT                          & 16                    & 36.5                     & 14.47                       & 0.421                    & -0.61                    & 0.686                            \\
		\multicolumn{1}{l}{} & Proposed BPT                      & 9                     & 20.78                    & 8.72                        & -0.118                   & -1.81                    & 0.245                            \\ \bottomrule
	\end{tabular}
\end{table}
}

We used the Welch's t-test for testing the differences in the correctness of answers, SUS and TLX scores between the \textit{Original BPT} and the \textit{Proposed BPT}. The Welch t-test is robust to different sample sizes even in the presence of deviations from normality \cite{kitchenhamrobust}. Table \ref{tab:welch} contains: the means for the average correction of answers, the SUS and the TLX scores; the difference between the original and new BPT means; the 95\% confidence interval of the difference; \textit{t}, \textit{df} and p-values.
{\footnotesize
\begin{table}[ht]
	\centering
	\caption{Welch's t-test scores}
	\label{tab:welch}
	\begin{tabular}{@{}lrrrrrrrr@{}}
		\toprule
		& \multicolumn{1}{c}{\begin{tabular}[c]{@{}c@{}}Orig\\BPT \\ mean\end{tabular}} & \multicolumn{1}{c}{\begin{tabular}[c]{@{}c@{}}Prop\\BPT\\ mean\end{tabular}} & \multicolumn{1}{c}{Diff} & \multicolumn{1}{c}{\begin{tabular}[c]{@{}c@{}}95\%\\ Dif. CI\\ Lower\end{tabular}} & \multicolumn{1}{c}{\begin{tabular}[c]{@{}c@{}}95\%\\ Dif. CI\\ Upper\end{tabular}} & \multicolumn{1}{c}{t} & \multicolumn{1}{c}{df} & \multicolumn{1}{c}{p-value} \\ \midrule
		Corr & 0.51 & 0.89 & -0.38 & -0.71 & -0.04 & -2.402 & 12.87 & 0.032\\
		SUS & 42.25                                                                   & 64.78                                                                      & -22.53                          & -31.83                                                                        & -13.22                                                                        & -5.07                 & 19.01                  & 0.000                       \\
		TLX & 36.50                                                                   & 20.78                                                                      & 15.72                          & 6.12                                                                          & 25.32                                                                         & 3.39                  & 22.80                  & 0.003                       \\ \bottomrule
	\end{tabular}
\end{table}
}
We hypothesised that the proposed version of BPT would lead to more correct interpretations of models, have a higher usability rating when compared to the original version and require a lower cognitive effort to be understood in its usage. 

\begin{figure*}
  	\begin{minipage}{0.33\textwidth}
 		\centering
\includegraphics[scale=0.33, trim = 3cm 12.5cm 2.6cm 2.7cm, clip]{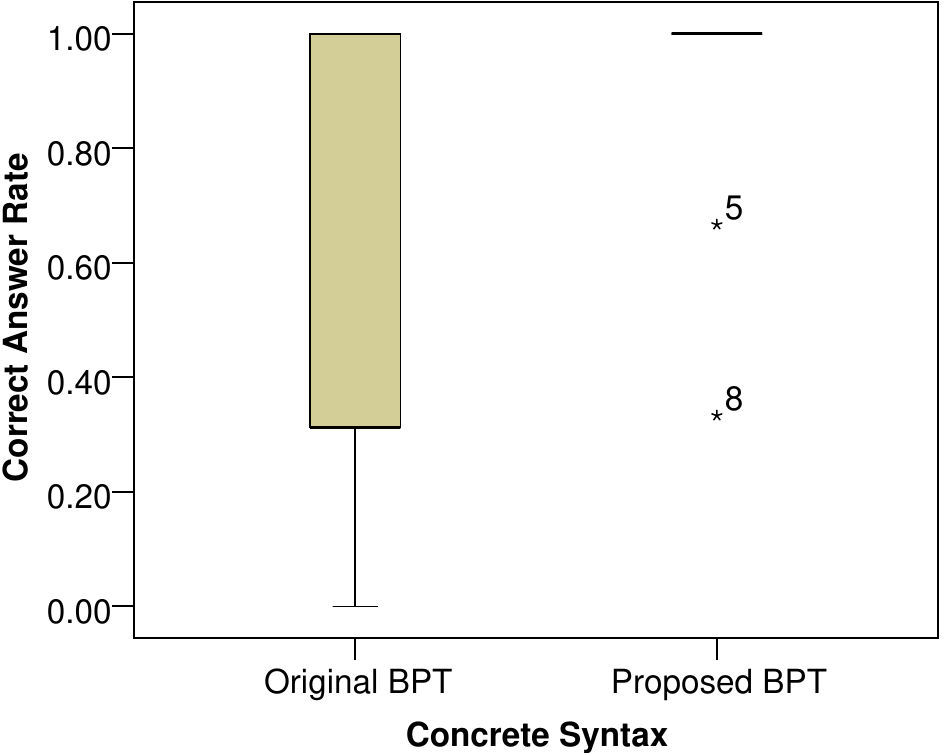}
 		\caption{TLX distribution}
		\label{fig:correctbox}
 	\end{minipage}\hfill
 	\centering
 	\begin{minipage}{0.33\textwidth}
 		\centering
\includegraphics[scale=0.33, trim = 3cm 12.5cm 2.6cm 2.7cm, clip]{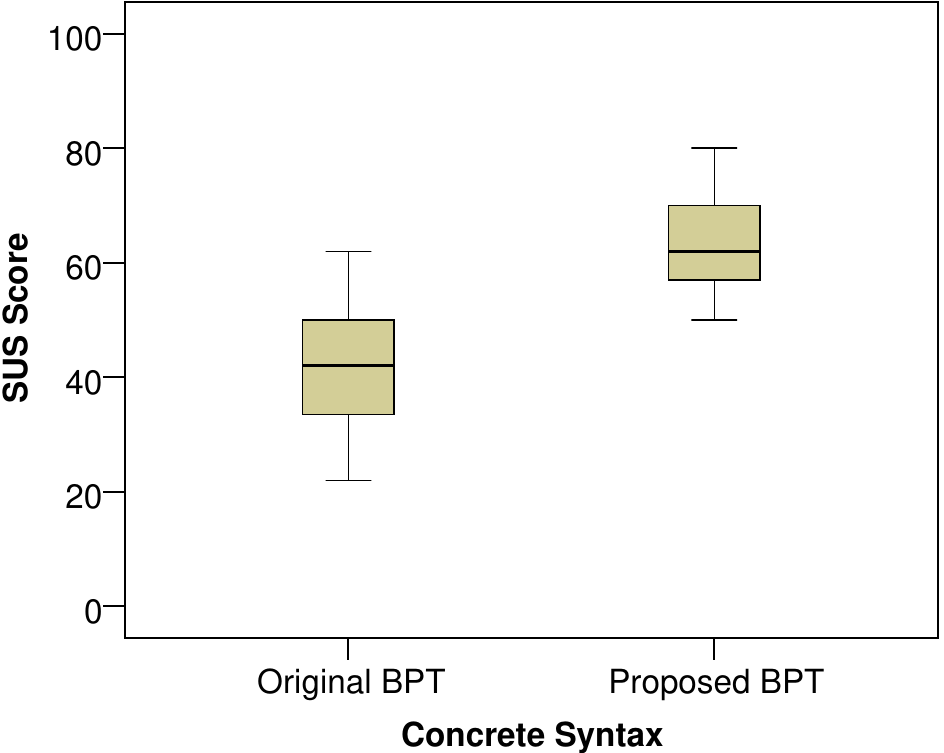}
 		\caption{SUS distribution} 		
		\label{fig:susbox}
 	\end{minipage}\hfill
 	\begin{minipage}{0.33\textwidth}
 		\centering
\includegraphics[scale=0.33, trim = 3cm 12.5cm 2.6cm 2.7cm, clip]{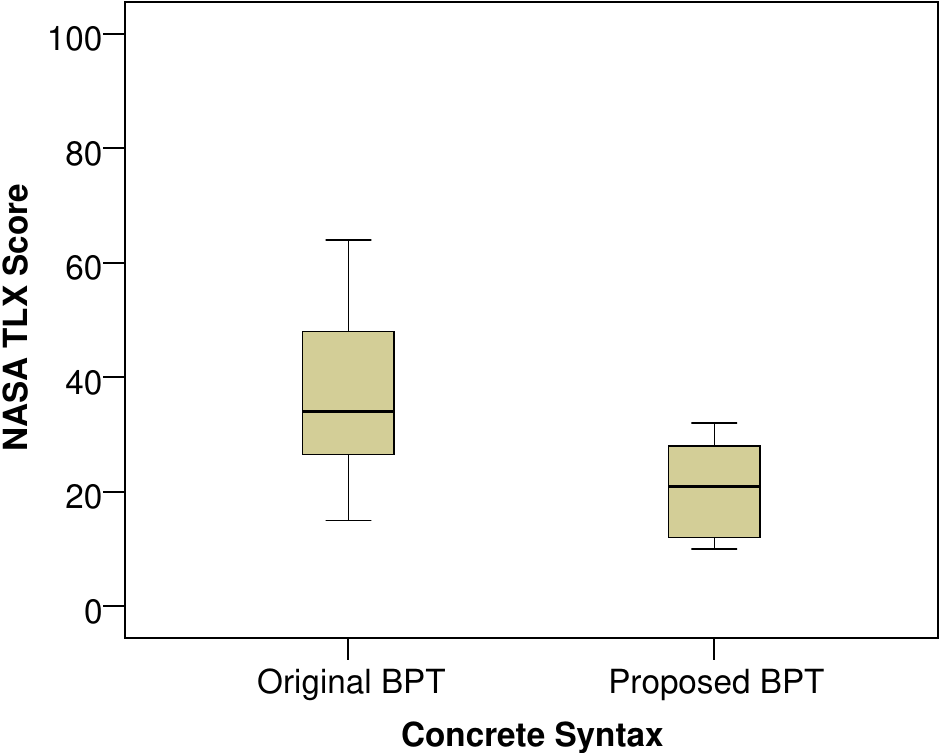}
 		\caption{TLX distribution}
		\label{fig:tlxbox}
 	\end{minipage}

 \end{figure*}

Participants using the \textit{proposed BPT} provided more correct answers, with a statistically significant improvement of 0.375. This supports the hypothesis that the \textit{proposed BPT} is easier to interpret than the original BPT.
Participants using the \textit{proposed BPT} gave it a higher score, with a statistically significant improvement of 22.53, supporting the hypothesis that the \textit{proposed BPT} leads to improved usability.
Finally, participants using the \textit{proposed BPT} reported a statistically significant lower NASA TLX score, with less 15.72 points, supporting the hypothesis that the proposed BPT requires a lower cognitive effort to be understood. Figures \ref{fig:correctbox}, \ref{fig:susbox} and \ref{fig:tlxbox} present the distributions of correctness, SUS and TLX, respectively.

%On  average, users who tested the original version of BPT gave it a SUS score of 42.25 and a TLX score of 36.50, while users who tested the new version of BPT gave it a SUS score of 64.78 and a TLX score of 20.78. The 95\% confidence interval of the difference for the effect of the new BPT on the SUS score is between -31.83 and -13.22 and on the TLX score it is between 6.12 and 25.32. As such, these results support our hypothesis.\par

\vspace{-0.2 cm}
\section{Discussion}
\label{sec:discussion}
\vspace{-0.1 cm}
\subsection{Evaluation of results}
Overall, the \textit{proposed BPT} significantly outperforms the \textit{original BPT} in terms of its actual and perceived usability and, therefore, has the potential for improving the developer experience with it.

\textbf{RQ1. Is the \textit{original BPT} semantically opaque?} Yes. The \textit{original BPT} has a mean hit rate well below the ISO standards requirements for symbol recognisability. This is very common in software design languages \cite{moody2009physics}. Along with the other shortcomings of the original BPT, identified through interviews with practitioners, the PoN evaluation and the scarce actual usage of BPT in projects, this observation suggests that improving BPT has the potential to improve developer and other stakeholders' experience by making BPT symbols easier to recognize and remember.

\textbf{RQ2. Can participants unfamiliar with BPT design more semantically transparent symbols for BPT than the original?} Yes. Asking participants unfamiliar with BPT to propose alternative representations for the BPT constructs has allowed obtaining more semantically transparent alternatives to BPT. Both the \textit{stereotype} and the \textit{prototype} alternatives have achieved a significantly better semantic transparency. These results reinforce others  where notations produced by novices consistently outperform those produced by experts, in terms of symbol recognizability \cite{caire2013visual,Santos2018ER}. This suggests that symbolization experiments such as ours are a viable way of developing better concrete syntaxes. A relevant difference from \cite{caire2013visual} is that instead of producing a PoN-powered alternative for a new concrete syntax \textit{a priori}, we proposed the improved BPT notation \textit{after} analyzing the results from the symbolization experiments. The proposed syntax was inspired by the alternatives previously produced (particularly the \textit{prototype BPT}), combining them it with other concerns, such as the overall coherence of the concrete syntax and how it relates to other existing DSLs in OutSystems.

\textbf{RQ3. Which concrete syntax (\textit{original}, \textit{stereotype}, \textit{prototype}, \textit{proposed}) is more semantically transparent?} The \textit{Prototype BPT} is the alternative with the best semantic transparency. However, its difference to the \textit{proposed BPT} and \textit{stereotype BPT} is not statistically significant. The three alternatives are significantly better than the baseline \textit{original BPT}. Again, this is somewhat similar to what was observed in other evaluations (see section \ref{sec:relatedWork}).

\textbf{RQ4. Which concrete syntax (\textit{original}, \textit{proposed}) leads to a better understandability of BPT models in the context of interpretation tasks?} The \textit{proposed BPT} lead to more correct interpretations in the model interpretation experiment. There were noticeable improvements: understanding where a process starts; whether and if a process fails; and, concerning the nodes finishing process flows. The insights collected in this evaluation can be used for further iterations in BPT, in particular for those details that participants still struggled with (Table \ref{tab:res}). More importantly, the approach itself is reusable for other languages. With some variations, it has been applied to other languages (see section \ref{sec:relatedWork}).

\textbf{RQ5. Which concrete syntax (\textit{original}, \textit{proposed}) leads to a better understandability of BPT models, as perceived by practitioners, while performing model interpretation tasks?} The \textit{proposed BPT} obtained a significantly higher SUS score, denoting that the perceived usability has improved.

\textbf{RQ6. Which concrete syntax (\textit{original}, \textit{proposed}) leads to a lower cognitive effort, as perceived by practitioners, while performing BPT model interpretation tasks?} Consistently with the perception of increased usability, the perceived cognitive effort has significantly decreased with the \textit{proposed BPT}.

%\vspace{-0.3 cm}
\subsection{Implications for practice}
The method followed in this paper is applicable to other languages. Indeed, we have partially done so elsewhere \cite{Santos2018ER,Miranda2018EMAS}. As shown in Table \ref{tab:HitRate}, symbols created by our participants were more transparent than those created by language engineers for the \textit{original BPT}. This is consistent with findings in other contexts \cite{caire2013visual, Santos2018ER, Miranda2018EMAS}. We introduced an important variant. Rather than using students as subjects \cite{caire2013visual, Santos2018ER, Miranda2018EMAS}, we had OutSystems professional developers as subjects. They are experts in the target platform, although inexperienced with BPT. This has facilitated the creation of the \textit{proposed BPT} as a visual language that is consistent with the rest of the OutSystems platform. Having participants with the same profile as the intended end users, but with a fresh look on the language concepts so they were not influenced by the current syntax of the language being evolved helped to achieve better results than those achieved with ``less informed'' participants in the symbolization experiments \cite{Santos2018RE}. 

The results of the usability experiments stress the importance of having a one-to-one relationship between the concrete and semantic constructs. This corroborates the Physics of Notations \cite{moody2009physics}. 

As expected, the \textbf{TLX} score decreases as the \textbf{SUS} score increases. This suggests a negative correlation between language usability and the perceived cognitive effort using it.

Our participants could not identify several constructs, in the \textit{original BPT}. This was one of the problems related to symbol overload or deficit, reinforcing the need to have a one-to-one relationship between the semantic constructs and the concrete syntax.

The changes introduced in the \textit{proposed BPT}, when compared to the \textit{prototype BPT}, in order to preserve consistency with the platform resulted in a minor, statistically insignificant, decrease in transparency. The \textit{proposed BPT} has still a mean hit rate above the ISO threshold. This illustrates the importance of context in language engineering (e.g. the colour choices for termination symbols only make sense for the context of the OutSystems platform). Understanding what works for the actual end users was key.

\vspace{-0.3 cm}
\subsection{Threats to validity}
We need to consider potential validity threats  \cite{wohlin2012experimentation}. Population selection is a threat, as, due to resource constraints, all the participants used in the usability experiments were members of OutSystems (not part of the \textbf{BPT} development team). Ideally, there would also be representatives of business managers, as they are also stakeholders for BPT. Further research is required to assess the \textit{proposed BPT} with those stakeholders. Also, due to the strict time availability of the participants (as is common in these experiments), the usability experiments were limited to three BPT diagrams of varying complexity. While those diagrams were selected for being as representative as possible of BPT, there is always the potential for increasing the external validity of these results by performing replications of this evaluation with different BPT diagrams.
\vspace{-0.1 cm}
\section{Related work}
\label{sec:relatedWork}

%\chapter{Related work}
%\label{cha:related}

%\section{Using Physics of Notations to evaluate BPMN 2.0}
Moody \textit{et al.} evaluated the \textit{i*} concrete syntax using PoN and proposed a new symbol set for it \cite{moody2010visual}. Caire \textit{et al.} compared Moody's proposed concrete syntax with alternatives produced by novices (a stereotype and a prototype concrete syntaxes) and the standard \textit{i*} concrete syntax \cite{caire2013visual}. We adapted Caire's protocol. Instead of having previously defined a PoN-informed concrete syntax for BPT, we also used the concrete syntaxes proposed through a symbolization experiment as input for the development of the \textit{proposed BPT}.

PoN was used to evaluate and identify improvement opportunities for several modelling languages, such as BPMN 2.0 \cite{genon2011analysing}, Use Case Maps \cite{genon2010analysing}, WebML  \cite{granada2017analysing}, and misuse cases \cite{saleh2015scientific}. In general, these studies reached conclusions similar to those advanced by Moody concerning the challenges in most visual notations, including UML, from a PoN perspective \cite{moody2009physics}.

Matulevičius \textit{et al.} used interviews, models creation, and evaluation of those models and the modelling language for assessing the \textit{i*} and KAOS modelling languages \cite{matulevivcius2007comparing} and found clarity problems in those languages semantics definition.
\vspace{-0.2 cm}
\section{Conclusions and future work}
\label{sec:conclusion}

% - Context;
% - Objectives;
% - Results;
% - Difficulties.

While running into maintenance costs, and having identified the need for improving the usability of the commercial business process modelling language (BPT) at OutSystems, we have designed and put forward a systematic process to identify and fix usability issues.

%% uncommon? lack of literature with this kind of reports?

We identified issues on syntactic and semantic constructs, and proposed a new notation. After evolving BPT within OutSystem's development environment, we applied the evaluation process to the proposed BPT and observed a significant increase in usability.

The comparison analysis between the original and the proposed version of the BPT confirmed that the process is effective and that the new notation has a higher usability rating. We could also conclude that: Semantic transparency has a large impact on usability; users create more semantic transparent symbols than language engineers (which goes in line with what Caire \textit{et al.} concluded \cite{caire2013visual}); it is extremely important to have a one-to-one relationship between the concrete syntax and the semantic constructs.

To be generalizable, as future work, the  evaluation process proposed in this work can be expanded with further techniques and also modified to apply to textual languages. The evaluation process identifies issues with a language's concrete and semantic constructs. However, it only provides methods to improve the language's concrete syntax. The process should be further expanded with techniques that help language engineers design semantic constructs from the ground up that answer the needs of the users while having a high usability rating. This could be achieved with the addition of techniques from Requirements Engineering, but further research is required. 
The proposed usability evaluation process should also make more use of the eye-tracking data. The current process only uses the eye-tracker to manually revisit recordings of the usability tests and to identify possible reasons for wrong answers. However, there are several metrics that can be extracted from the eye-tracking results that provide objective information about the tester (states of confusion, being lost in the interface, etc.). This, added to the SUS score provides a more accurate overview of the language's usability. During the analysis of the comparison between the original BPT notation and the new one, we noted that there appears to be a negative correlation between a language's SUS and NASA TLX. This should be further researched.

\section*{Acknowledgments}
The authors would like to thank  NOVA LINCS Research Laboratory (Grant: FCT/MCTES PEst UID/ CEC/04516/2013) and DSML4MAS Project (Grant: FCT/MCTES TUBITAK/0008/2014).

\bibliographystyle{ACM-Reference-Format}
\bibliography{references}

\end{document}